\documentclass[conference]{IEEEtran}

\ifCLASSINFOpdf
\else
\fi
\usepackage{graphicx}
\usepackage{subfigure}
\usepackage{amsmath}

\begin{document}
%
\title{Quasi-stationary Slice Detection-Based Robust Respiration Rate Estimation under Large-scale Random Body Movement}
\author{\IEEEauthorblockN{Chendong Xu$^1$, Shuai Yao$^1$, Haoying Bao$^2$, Chiyuan Ma$^{\star2}$, Qisong Wu$^{\star1}$}
\IEEEauthorblockA{$^1$Key Laboratory of Underwater Acoustic Signal Processing of Ministry of Education, \\
Southeast University, Nanjing 210096, China\\
$^2$Department of Neurosurgery, Eastern Theater Hospital, School of Medicine, \\
Southeast University, Nanjing 210002, China \\
Email: 230228233@seu.edu.cn, yaoshuaiseu@seu.edu.cn, Haoying.bao@sina.com, \\
Machiyuan75@sina.com, qisong.wu@seu.edu.cn}
}



%


\maketitle

\begin{abstract}

Radar-based non-contact respiration rate (RR) measurement has become increasingly popular due to its convenience, non-intrusiveness, and low cost. However, it is still quite challenging to accurately acquire vital signs estimation in complex measurement scenarios with large-scale random body movements (RBM), particularly for RR estimation due to strong low-frequency interferences. To cope with the RBM challenge in RR estimation, we propose a novel two-stage RR estimation scheme involving detecting the portion of signals, called as quasi-stationary slices, exhibiting the quasi-stationary pattern. At the detection stage, an enhanced deep neural network framework incorporating the dynamic snake convolution is exploited to detect the quasi-stationary slices in the micro-Doppler spectra. At the estimation stage, we mitigate RBM interferences and achieve accurate RR estimation by only using the portion of ridges consistent with the location of detected quasi-stationary slice. Extensive experimental results demonstrate that our proposed scheme can accurately detect quasi-stationary slices under normal scenarios with large-scale RBM, thereby reducing the error of subsequent RR estimation.

\end{abstract}


\IEEEpeerreviewmaketitle

\section{Introduction}
Respiration rate (RR), serving as a crucial health assessment indicator, holds the potential to reveal various cardiopulmonary conditions \cite{David}. In the context of illnesses such as COVID-19, RR serves as a metric to identify pneumonia in patients \cite{Massaroni}. Conventional techniques predominantly employ wearable devices for RR monitoring. Nevertheless, these devices prove impractical for individuals with skin conditions and may cause discomfort, particularly in sensitive populations like children, potentially impacting their breathing. Radar-based non-contact techniques for measuring respiration rate (RR) have witnessed a surge in popularity, attributed to their convenience, non-intrusiveness, and cost-effectiveness, and thus this subject has captured growing interest among researchers. 

Radar-based vital signs detection relies on sensing minute physiological movements ranging from several millimeters to several centimeters in the chest wall. However, the presence of random body movements (RBM), characterized by displacements comparable to or larger than those caused by vital signs in the chest wall, pose significant interferences capable of compromising the signals of interest. Consequently, this can lead to a considerable degradation in the accuracy of vital signs estimation, particularly in the estimation of RR. It is because the spectrum of RBM often manifests as strong interference in the low-frequency range, frequently overlapping significantly with the frequency range associated with RR \cite{Rahman}. While radar-based vital signs monitoring technology has made considerable progress, accurately estimating the RR of human subjects with substantial RBM remains a significant challenge.

Numerous endeavors have been dedicated to addressing this challenge \cite{Shang}-\cite{Gu}.  One direct approach involves identifying data blocks affected by RBM through energy or feature threshold detection techniques, and then achieves RR estimation by discarding the RBM-affected data blocks \cite{Shang}. A blind source separation algorithm was introduced to estimate vital signs by isolating weak vital sign signals from strong RBM interferences \cite{Jutten}. The spectrum characteristics of vital sign signals under 1-D body motion were analyzed and RR measurement was achieved using continuous wave (CW) Doppler radar after motion direction detection \cite{Tu}. Hidden Markov Model (HMM) techniques were employed to measure RR and heart rate (HR) in the presence of large-scale RBM \cite{Wu}. The different available portions of RR and HR signals were identified and quantified by several traditional classifiers, thereby detecting and excluding the signals with large disturbances caused by inevitable body movements \cite{XieSQD}. A two-stream scheme estimated the RR and HR respectively by capturing the local and global features to address the challenges from nonlinear composition between respiration and heartbeat signals, as well as body movements \cite{DeepVS}. To mitigate RBM effects, studies have explored the use of two Doppler radar systems, considering different body orientations for RBM cancellation \cite{Li}. Additionally, self and mutual injection-locking techniques have been applied using two radar systems to cancel out RBM \cite{Wang}. Moreover, some collaborative camera-radar systems were devised for non-contact vital sign monitoring, utilizing auxiliary cameras to mitigate the effects of RBM interferences or distinguish different individuals \cite{Gu, VitalHub}.  

In this paper, a novel two-stage RR estimation scheme involving quasi-stationary slice detection in the deep neural network (DNN) framework is developed in the presence of large-scale RBMs. The short-time Fourier transform (STFT) method is first performed to acquire the micro-Doppler spectrum, in which the quasi-stationary slices exhibit periodic characteristics of chest displacement. Motivated by the unique periodicity of quasi-stationary slices, our proposed YOLOv8-based \cite{YOLOv8} scheme, named DSC-YOLOv8, achieves the RR estimation in two stages by detecting and employing the quasi-stationary slices. Compared to previous versions, YOLOv8 is designed to be more accurate and faster, making it suitable for practical applications across various fields, and which is the main motivation for developing our DNN framework primarily based on YOLOv8.  
At the detection stage, a DNN framework is carefully devised to find out the locations of quasi-stationary slices. In particular, a dynamic snake convolution (DSC) \cite{DSC} is introduced to enhance the YOLOv8's capability of modeling tubular structures in the micro-Doppler spectrum. At the estimation stage, the ridges are first extracted from the micro-Doppler spectrum by directly detecting the maximum magnitude in the frequency direction. Then, to mitigate RBM interferences, the portions of ridges corresponding to the detected quasi-stationary slices, called as truncated ridges, are retained and the remaining slices containing RBM are discarded. 
Therefore, the accurate RR can be estimated by only using the truncated ridges without RBM. Extensive experiments demonstrate that the proposed DSC-YOLOv8 greatly improves the accuracy of quasi-stationary slice detection compared to state-of-the-art (SOTA) methods and subsequently reduces the error of RR estimation under large scale RBM.

The rest of this paper is organized as follows. Section \uppercase\expandafter{\romannumeral2} introduces our proposed scheme. Section \uppercase\expandafter{\romannumeral3} presents the experimental results. Section \uppercase\expandafter{\romannumeral4} summarizes the full paper.

\section{Proposed Method}
Under quasi-stationary positions, the chest displacement caused by breathing varies regularly and stably, so the corresponding slices in micro-Doppler spectrum, named the quasi-stationary slices, should exhibit periodic characteristics. Inspired by this unique property, the accurate RR during this period can be estimated by detecting and employing these slices. In this section, we first introduce the signal model and data preprocessing in Subsection A. Then, we devise a DNN framework for detecting quasi-stationary slices in the spectrum and illustrate how to estimate RR after the detection stage in Subsection B.

\begin{figure}[t]
	\centering 
		\subfigure{
		\label{freq1}
		\includegraphics[width=35mm,height=26mm]{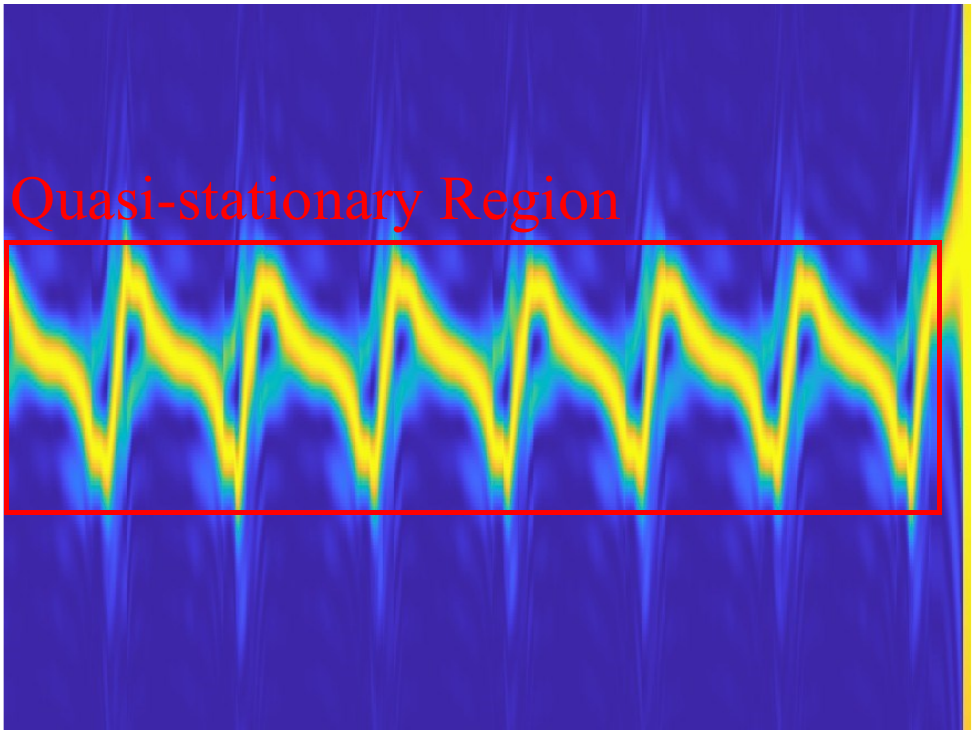}}
		\subfigure{
		\label{freq2}
		\includegraphics[width=35mm,height=26mm]{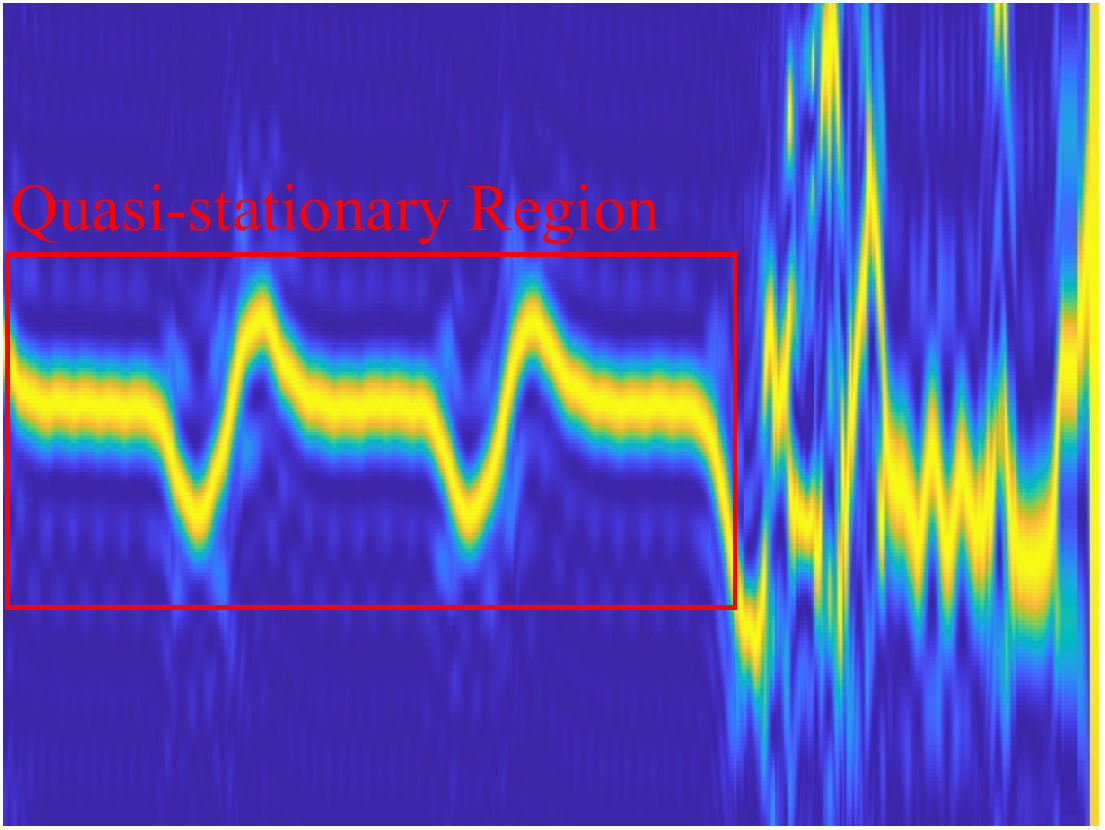}}
	\caption{Micro-Doppler spectra of vital sign signal and annotations of quasi-stationary slices.}
	\label{freq}
\end{figure}

\subsection{Signal Model and Data Preprocessing}
Taking a simple sinuous continuous waveform radar for example, the transmitted waveform can be expressed by,
\begin{equation}
	{s_{TX}} = \cos (2\pi {f_0}t + \phi (t)) \text{,}
\end{equation}
where $f_0$ is the carrier frequency and $\phi (t)$ is the phase noise. Suppose that the distance between the chest and the radar is $d_0$ and the displacement of the chest is $r(t)$. The real-time distance between the radar and the chest can be expressed as
\begin{equation}
	\label{eq2}
	d(t) = {d_0} + r(t) \text{.}
\end{equation}
The displacements caused by respiration and heartbeat are often modeled as sinusoidal signals \cite{Shang, QW23GRSL, QW23TIM}, so the chest displacement $ r(t)$ can be approximately formulated as
\begin{equation}
	\label{eq3}
	r(t) \approx {A_r}\sin (2\pi {f_r}t) + {A_h}\sin (2\pi {f_h}t) \text{,}
\end{equation}
where $A_r$ and $A_h$ are the amplitude of chest displacement caused by respiration and heartbeat respectively, and $f_r$ and $f_h$ are the frequencies of respiration and heartbeat respectively. Therefore, the reflected signal can be expressed as
\begin{equation}
	{s_{RX}} = \cos (2\pi {f_0}(t - \tau ) + \phi (t - \tau )) \text{,}
\end{equation}
where $\tau = 2d(t)/c$ is the round-trip time delay and $c$ is the speed of electromagnetic wave. The mixer with the reference waveform is used, and the low-pass filter is then employed to acquire the beat signal $s_b(t)$. The STFT is subsequently employed to generate micro-Doppler spectrum. The fundamental idea of STFT is to apply the Fourier transform to local segments of the signal obtained by the sliding window function $w(m)$. The STFT of a discrete beat signal $s_b(n)$ with $n=1,\cdots,N$, which denote the samples of $s_b(t)$,  can be  written as
\begin{equation}
	STFT(n,k) = \sum\limits_{m = 0}^{M - 1} {s_b(n + m)w(m){e^{ - j\frac{{2\pi }}{N}mk}}}
\end{equation}
where $n$ and $k$ represent time instant and frequency respectively, and $M$ represents the width of the window.

\begin{figure*}
	\centering 
	\includegraphics[width=160mm,height=35mm]{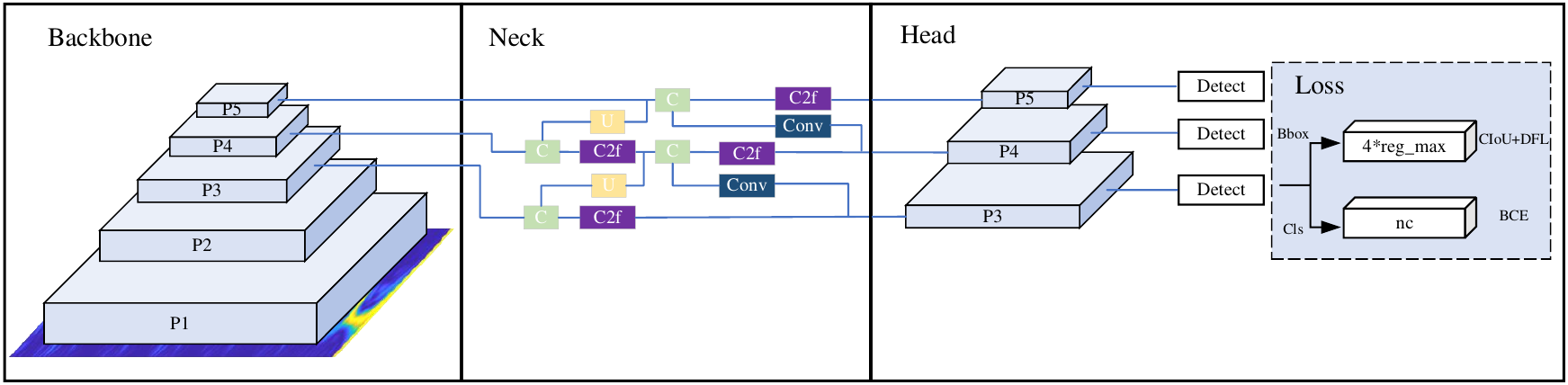}
	\caption{Detailed architecture of YOLOV8. "C", "U", "C2f" and "Conv" denote the Concat operation, upsampling operation, C2f module in YOLOv8 and convolutional layer respectively.}
	\label{Net}
\end{figure*}

\begin{figure*}
	\centering 
	\includegraphics[scale=0.45]{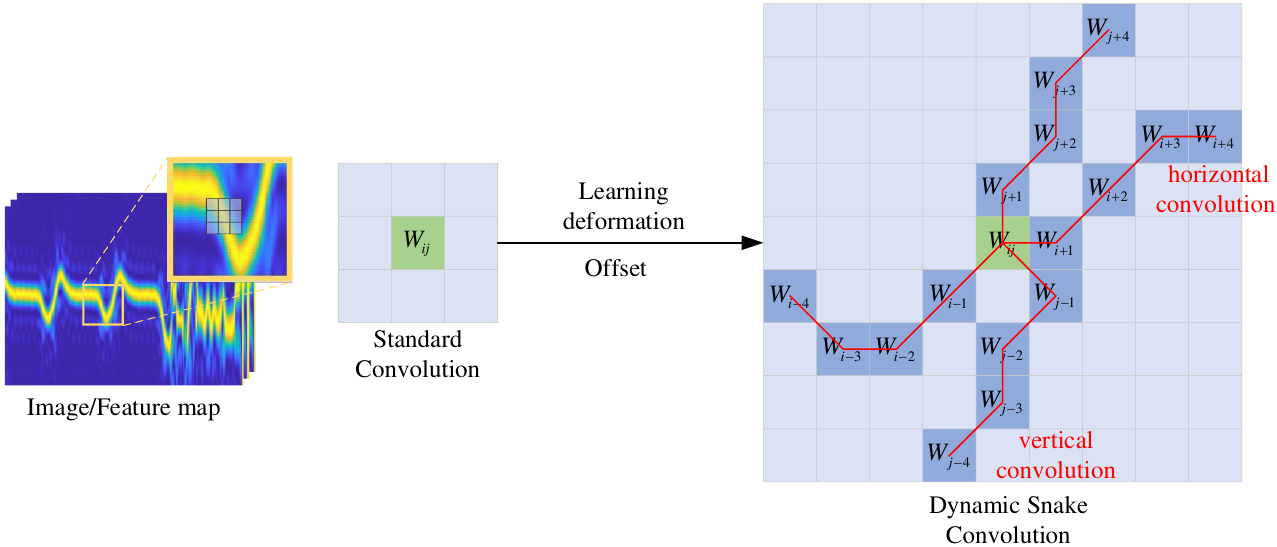}
	\caption{Deformation process of dynamic snake convolution (DSC). DSC adaptively captures the features with the deformable convolutional kernels in horizontal/vertical direction. The learned deformation offset forces the DSC to focus on the slender and tortuous structure.}
	\label{DSC}
\end{figure*}

In the absence of RBM, from Eq. \ref{eq2} and Eq. \ref{eq3}, it is readily derived that the distance between the chest and the radar is a periodic signal. The periodic variations of the quasi-stationary slices in the red solid box in Fig. \ref{freq} also prove this.
However, RBM with more powerful energy in other slices can destroy this signal pattern, mask vital sign signals and lead to erroneous RR estimation results. If the entire signal is directly discarded just because some slices contain RBM, it will result in long-term unusable period and data waste. Considering that quasi-stationary slices without RBM interferences can be directly used for accurate RR estimation, we estimate RR by detecting them instead of using the entire signal. Before detection, we first annotate the spectrum, and slices with periodic micro-Doppler spectrum are labeled as Quasi-stationary Slice.

\subsection{DSC-YOLOv8}
We choose the YOLOv8 as our baseline method due to its high accuracy and speed in quasi-stationary slice detection.
Fig. \ref{Net} shows the three key components of YOLOv8: backbone, neck and head. YOLOv8 uses an improved CSPDarkNet with a richer gradient flow as the network backbone to extract image features. The neck and head are used to predict the target category and location. Since the classification and localization tasks in object detection focus on different areas, using a shared detection head will cause spatial misalignment problems \cite{Decoupled}. Therefore, the shared head for classification and localization is replaced by a decoupled head. YOLOv8 uses a loss function that combines classification loss and localization loss to optimize the network. The total loss is defined as
\begin{equation}
	L = \alpha {L_{CIOU}} + \beta {L_{Cls}} + (1 - \alpha  - \beta ){L_{DFL}} \text{,}
\end{equation}
where ${L_{CIOU}}$ and ${L_{DFL}}$ are used for bounding box regression, ${L_{Cls}}$ is used to calculate category loss, and $\alpha$ and $\beta$ are used to adjust the importance of different losses.

Although YOLOv8 has powerful feature extraction capabilities, it still lacks the ability to model the geometric transformations due to the limitation of convolutional neural networks with fixed sampling locations. Unlike objects with rectangular shape, micro-Doppler spectrum exhibits slender and tubular structure. To enhance the feature capture ability of YOLOv8 for this special structure, we introduce a DSC with adaptive sampling locations.
The standard 3*3 2D convolution kernel with 9 sampling locations can be expressed as
\begin{equation}
	W = \left\{ {\left( {{x_o} - 1,{y_o} - 1} \right), \cdots \left( {{x_o},{y_o}} \right), \cdots ,\left( {{x_o} + 1,{y_o} + 1} \right)} \right\} \text{,}
\end{equation}
where  ${({x_o},{y_o})}$ represents the center location of the convolution kernel. For each location $p_0$ in the feature map, the standard convolution process can be expressed as
\begin{equation}
	Y\left( {{p_0}} \right) = \sum\limits_{{p_n} \in W} {W({p_n}) \cdot X({p_0} + {p_n})} \text{.}
\end{equation}

\begin{figure*}
	\centering 
	\includegraphics[scale=0.45]{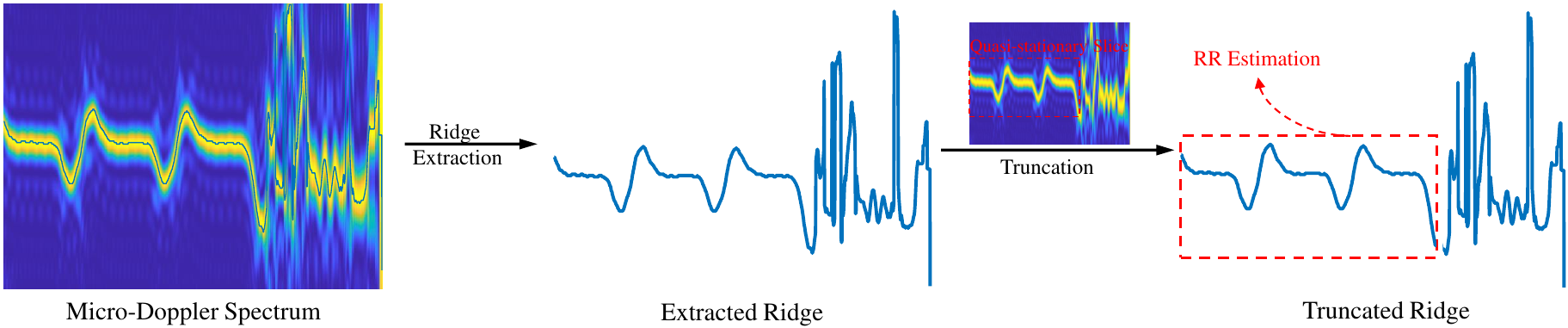}
	\caption{Ridge extraction and truncation process. The Ridge is first extracted from the spectrum and then truncated according to the detected quasi-stationary slice. The accurate RR can be estimated by only using the truncated ridge.}
	\label{Ridge}
\end{figure*}

\begin{figure}
	\centering 
	\subfigure{
		\label{age}
	\includegraphics[scale=0.3]{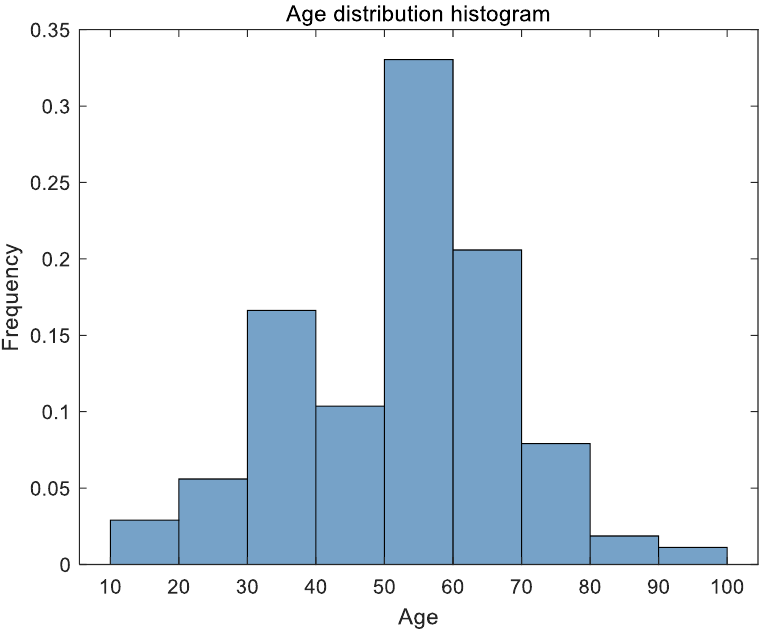}}
	\subfigure{
	\label{RR}
	\includegraphics[scale=0.3]{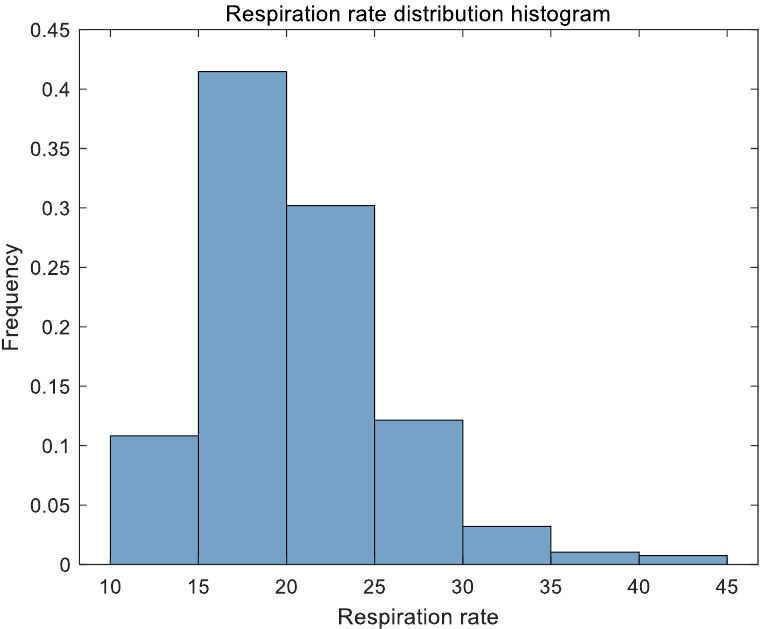}}
\caption{Detailed data distribution histogram.}
\label{histgram}
\end{figure}

\begin{figure}
	\centering 
	\includegraphics[scale=0.3]{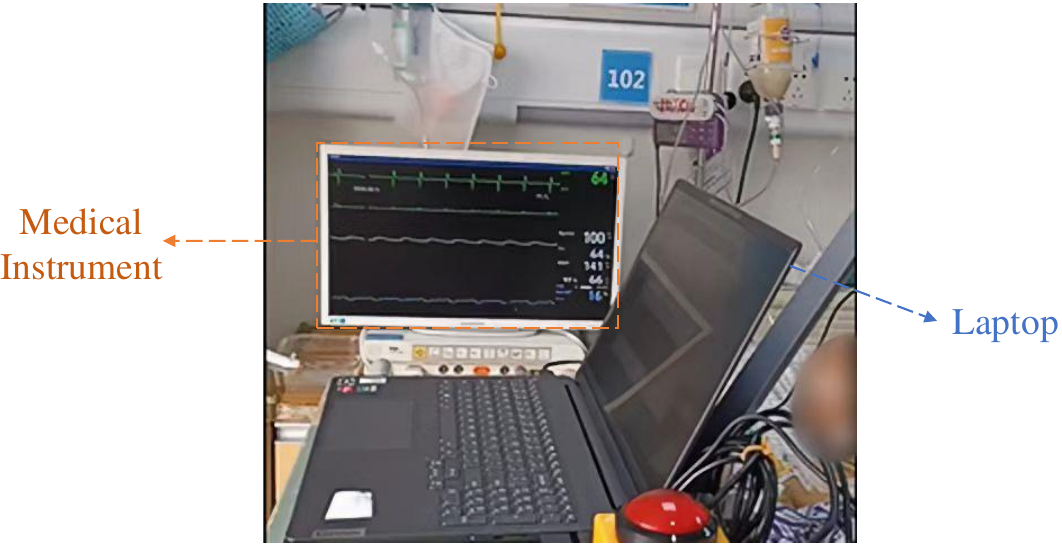}
	\caption{Data collection scenario.}
	\label{scenario}
\end{figure}

To ensure that the receptive field adaptively focuses on the slender and tortuous local structures, DSC introduces deformation offset $\Delta$ for sampling process and determines each sampling location according to the sum of offsets of all previous sampling locations. This sampling strategy also ensures that the receptive field is continuous and does not deviate too far from the target.
The kernel of DSC can be defined as ${W_{ij\pm c}} = ({x_{ij\pm c}},{y_{ij\pm c}})$, where $(x_{ij},y_{ij})$ represents the center location, and $c = \left\{ {0,1,2,3,4} \right\}$ represents the distance from the center location. As shown in Fig. \ref{DSC}, DSC captures the features of micro-Doppler spectrum in two directions through two convolutions. The convolution in the horizontal direction can be expressed as
\begin{equation}
	{W_{i \pm c}} = \left\{ {\begin{array}{*{20}{c}}
			{({x_{i + c}},{y_{i + c}}) = ({x_i} + c,{y_i} + \alpha )} \text{,}\\
			{({x_{i - c}},{y_{i - c}}) = ({x_i} - c,{y_i} + \beta )} \text{,}
	\end{array}} \right.
\end{equation}
where $\alpha  = \sum\nolimits_i^{i + c} {\Delta y} $, $\beta  = \sum\nolimits_{i - c}^i {\Delta y} $. The convolution in the vertical direction can be expressed as
\begin{equation}
	{W_{j \pm c}} = \left\{ {\begin{array}{*{20}{c}}
			{({x_{j + c}},{y_{j + c}}) = ({x_j} + \gamma ,{y_j} + c)} \text{,}\\
			{({x_{j - c}},{y_{j - c}}) = ({x_j} + \psi ,{y_j} - c)} \text{,}
	\end{array}} \right.
\end{equation}
where $\gamma  = \sum\nolimits_j^{j + c} {\Delta y} $, $\psi  = \sum\nolimits_{j - c}^j {\Delta y} $.  Since DSC can accurately capture the slender and tortuous local features better than standard convolution, we carefully devise a DNN framework for detecting quasi-stationary slices by replacing the standard convolution in YOLOv8 with DSC. Unlike the traditional YOLOv8 with the fixed receptive field, our framework can better perceive the key features with the help of adaptive receptive field.

\begin{figure*}
	\centering 
	\includegraphics[scale=0.4]{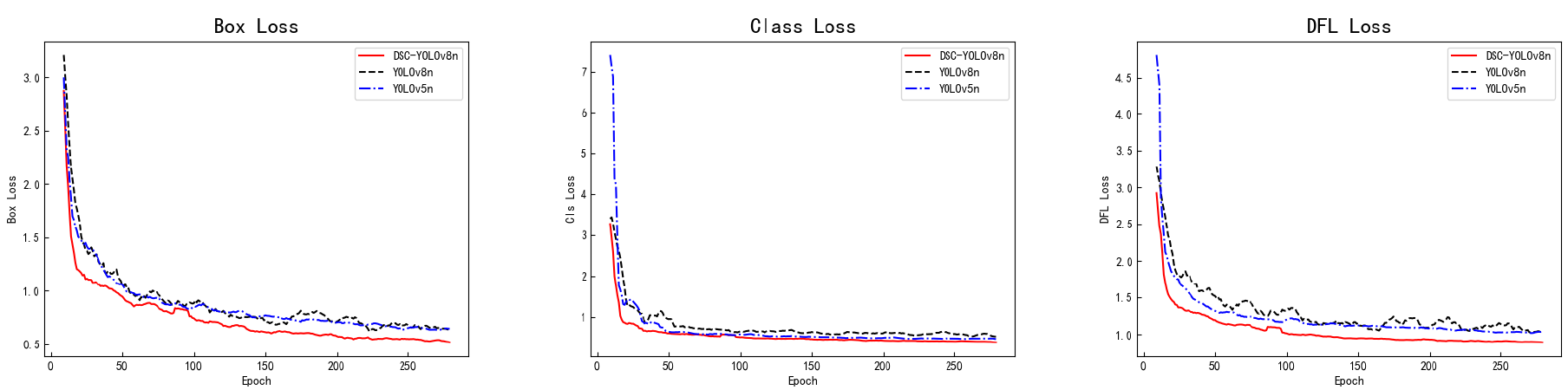}
	\caption{Comparison of different losses. The loss of DSC-YOLOv8 decreases fastest and reaches the lowest level.}
	\label{loss}
\end{figure*}

\begin{figure*}
	\centering 
	\includegraphics[scale=0.45]{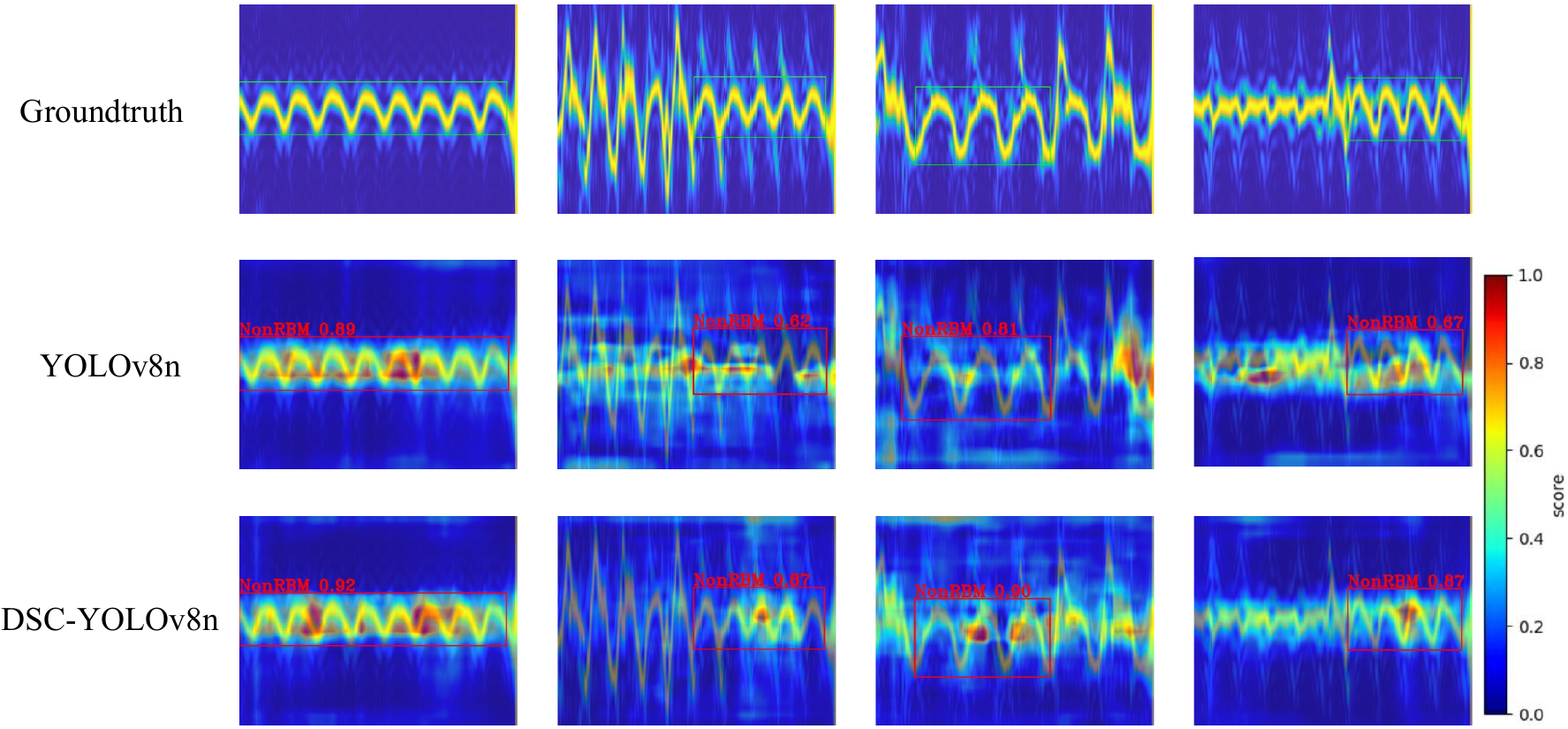}
	\caption{Heatmap comparison of DSC-YOLOv8 and YOLOv8. Regions with higher scores contribute more to network detection results.}
	\label{heatmap}
\end{figure*}

The ridge of the micro-Doppler spectrum is defined as the location where the magnitude reaches its local maximum along the frequency direction.
Therefore, we directly extract the ridge from the spectrum $STFT(n,k)$ by detecting the maximum magnitude in the frequency direction for every time instant $n$ \cite{Liu}. The direct ridge extraction can be formulated as
\begin{equation}
	R(n) = \mathop {\max }\limits_k STFT(n,k) \text{.}
\end{equation}
As shown in Fig. \ref*{Ridge}, the extracted ridge contains the same trend of the micro-Doppler spectrum.
To mitigate the RBM interferences, we first truncate all the ridges according to the locations of the detected quasi-stationary slices. Then, the RR of the quasi-stationary slices can be estimated by calculating the frequency of the truncated ridges without RBM.

\begin{table}
	\centering
	\label{radar} 
	\caption{KEY PARAMETERS OF THE FMCW RADAR IN EXPERIMENTS}
	\begin{tabular}{l|l}
		\hline
		Parameters                  & Value  \\
		\hline
		Number of Transmit Antennas & 3      \\
		\hline
		Number of Receive Antennas  & 4      \\
		\hline
		Starting Frequency          & 60GHz      \\
		\hline
		Bandwidth                   & 4GHz      \\
		\hline
		ADC Sampling Rate           & 25000      \\
		\hline
		Chirp Duration              & 0.625ms      \\
		\hline
		Number of Chirps per Frame  & 1      \\
		\hline
		Frames per Second           & 50     \\
		\hline
		Range Resolution            & 3.75cm      \\
		\hline
	\end{tabular}
\end{table}

\subsection{Summary of the Proposed Respiration Rate Estimation Scheme}
The proposed scheme is summarized in Algorithm 1.
\begin{table}[h]
	\label{radar} 
	\begin{tabular}{p{8.5cm}} 
		\hline
		\textbf{Algorithm1} Summary of the Proposed Respiration Rate Estimation Scheme \\
		\hline
		\textbf{Input:}	The  beat signal of CW radar	$s_{b}(n)$			\\
		\textbf{Step1:} Perform STFT on the $s_{b}(n)$ and obtain the micro-Doppler spectrum $STFT(n,k)$. \\
		\textbf{Step2:} Train the enhanced YOLOv8 with annotated data to detect the quasi-stationary slices\\
		\textbf{Step3:} Extract the ridge from the micro-Doppler spectrum $STFT(n,k)$. Truncate the ridges according to the detected quasi-stationary slices. Estimate RR by using the truncated ridges.\\
		\hline
		\textbf{Output:} Robust RR estimation\\
		\hline
	\end{tabular}
\end{table}

\section{Experimental Results}
We collected 1,260 discontinuous sets of respiratory data from patients with different cardiopulmonary diseases at the Eastern Theater Hospital in Nanjing, China. The duration of each set was 20 seconds and converted into a single spectrum that does not overlap with other spectra. Patients of different ages and respiration rates were fully considered, and the detailed data distribution is shown in Fig. \ref{histgram}. Patients are not strictly required to remain still during data collection. The IWR 6843  frequency-modulated continuous wave radar manufactured by Texas Instrument was used to collect data. Its key parameters are reported in Table \uppercase\expandafter{\romannumeral1}. Meanwhile, as shown in Fig. \ref{scenario}, a medical instrument collects the real RR as reference values, and a laptop is used to store data. The DSC-YOLOv8 is implemented on an Ubuntu 16.04 server with Tesla A100 Graphics Processing Unit (GPU). The network is optimized by using stochastic gradient descent (SGD) with a learning rate of 0.001 and a momentum of 0.9. The input spectrum size is 875*656 and is scaled to 896*672.

\begin{table}
	\centering
	\label{comparison} 
	\caption{Performance comparison of the SOTA models}
	\begin{tabular}{{c|c|c|c|c}}
		\hline
		Model 		   & mAP0.5-0.95 		  & FPS   		   & FLOPs 			& Params \\
		\hline
		Faster RCNN    & 0.625                & 23.36          & 487.70 G		& 28.28M		  \\
		RetinaNet      & 0.642                & 58.14          & 115.82 G       & 36.33 M         \\
		CenterNet      & 0.652                & 72.33          & 79.03 G        & 32.67 M         \\
		YOLOv5n        & 0.764                & \textbf{131.58}   & 7.1 G          & 2.50 M          \\
		YOLOv8n        & 0.766                & 123.46		   & 8.1 G          & 3.00 M          \\
		\hline
		DSC-YOLOv8n    & \textbf{0.787}       & 93.46          & 8.2 G          & 3.27 M 		  \\
		\hline        
	\end{tabular}
\end{table}

\begin{table}[]
	\centering
	\label{YOLO} 
	\caption{Comparison of the respiration rate estimation}
	\begin{tabular}{c|cc}
		\hline
		Model			& Accuracy  & MAE 	\\
		\hline
		FFT				& 77.34		& 2.43	\\
		YOLOv5n     	& 78.90     & 2.23  \\
		YOLOv8n     	& 79.54     & 2.19  \\
		\hline
		DSC-YOLOv8n 	& \textbf{81.25}     & \textbf{2.13}	\\
		\hline
	\end{tabular}
\end{table}

\subsection{Comparison with SOTA Models}
In this section, we compare the performance of DSC-YOLOv8n, including mean average precision for Intersection over Union (IoU) increasing from 0.5 to 0.95 (mAP0.5-0.95), frames per second (FPS), floating point operations per second (FLOPs), and model parameters (Params), with SOTA algorithms. The metric mAP directly determines whether quasi-stationary sclices can be correctly located, thereby affecting the performance of subsequent RR estimation. As shown in Table \uppercase\expandafter{\romannumeral2}, DSC-YOLOv8n achieves the best performance with a slight loss of speed. Faster RCNN, as a two-stage object detection algorithm, has the highest number of parameters and computational complexity, yet resulting in the lowest speed and accuracy. In contrast, RetinaNet and CenterNet, which are one-stage object detection algorithms, have fewer parameters and lower computational complexity, resulting in faster speed and comparable accuracy compared to Faster RCNN. The networks of the YOLO family can achieve higher accuracy and speed compared to other networks, demonstrating the effectiveness of the YOLO architecture. Among them, YOLOv5n and YOLOv8n achieve similar accuracy, but YOLOv5n has the fastest speed. The proposed DSC-YOLOv8n model achieves the highest accuracy at 78.7\%, FPS at 93.46, and requires FLOPs and parameters similar to YOLOv8n. Overall, compared to other models, DSC-YOLOv8n performs well in terms of both accuracy and speed for NonRBM detection.

\subsection{Ablation Experiments}
To investigate the improvement of RR estimation by DSC-YOLOv8n, we conducted several ablation experiments. We estimated the RR using the truncated ridges without RBM and compared the RR estimation performance of DSC-YOLOv8n with YOLOv8n and YOLOv5n. Two performance metrics, accuracy and Mean absolute error (MAE), are employed in this study. An estimation is defined as accurate when the absolute error between the estimated RR and the reference RR is within the range of 3 beats per minute (bpm) for respiratory tasks.
At the same time, the performance of directly using FFT to estimate RR of the entire signal without detecting quasi-stationary slices was also compared. As shown in Table \uppercase\expandafter{\romannumeral3}, compared with directly estimating RR, the RR estimation accuracy of DSC-YOLOv8n is improved by 3.91\%, while the error is reduced by 0.3 bpm. This suggests that more accurate RR estimates can be obtained by detecting quasi-stationary slices. Meanwhile, compared with the other two detection networks, DSC-YOLOv8n has also greatly improved the performance of RR estimation.
In Fig. \ref{loss}, we compared three different loss functions used during training. ${L_{CIOU}}$, ${L_{Cls}}$ and ${L_{DFL}}$ are shown in the left, middle, and right figures respectively. We observed that DSC-YOLOv8n, with the introduction of DSC, exhibits a faster decline in all three losses and reaches a lower level. It indicates that DSC-YOLOv8n outperforms YOLOv8n. Additionally, the comparison of heatmaps between the predictions of YOLOv8n and DSC-YOLOv8n for some discontinuous spectra is presented in Fig. \ref{heatmap}. A heatmap is used to visualize the contribution of various parts of the spectrum to network predictions. It can be observed that YOLOv8n often focuses on the regions outside of the quasi-stationary slices. In contrast, DSC-YOLOv8n can more accurately capture slender tortuous features and predict with more confidence.

\section{Conclusion}
The paper introduces a novel two-stage vital signs estimation scheme that incorporates quasi-stationary slice detection within the deep neural network (DNN) framework to achieve precise respiration rate (RR) estimation in the presence of large-scale random body movements (RBMs). The enhanced DNN method was specifically designed to improve the performance of YOLOv8 in detecting quasi-stationary slices by introducing dynamic snake convolution (DSC) to effectively capture the slender tubular structures. By utilizing truncated ridges consistent with the locations of detected quasi-stationary slices, the proposed method successfully mitigates RBM interferences, leading to robust and accurate RR estimates. Extensive experiments have unequivocally demonstrated the superior performance of the proposed DSC-YOLOv8 approach. Furthermore, the two-stage vital signs estimation scheme holds promise for various application scenarios, including clinical diagnosis for patients and health assessment for sleeping individuals and drivers. In these contexts, DSC-YOLOv8 demonstrates its ability to robustly estimate RR and minimize the impact of RBM interferences.


\section*{Acknowledgment}

The work was supported by the National Natural Science Foundation of China under Grants No. 12374421 and 91938203, and in part by Key Laboratory of Underwater Acoustic Countermeasure Technology under Grant No. 2022JCJQLB03305, in part by National Key Laboratory of Science and Technology on Underwater Acoustic Antagonizing under Grant No. JCKY2023207CH01, in part by Fundamental Research Funds for the Central University under Grants No.2242023K30003 and 2242023K30004.



%

\end{document}